\documentclass[12pt]{article}

\usepackage{latexsym,amsfonts,amsmath,amssymb,pstricks,wasysym,stmaryrd,enumerate,epsfig}
\usepackage{mathrsfs}
\usepackage{theorem}

\numberwithin{equation}{section}
\newtheorem{definition}{Definition}[section]

\newtheorem{claim}[definition]{Claim}

\newcommand{\be}{\begin{equation}}
\newcommand{\ee}{\end{equation}}
\newcommand{\beu}{\begin{equation*}}
\newcommand{\eeu}{\end{equation*}}
\newcommand{\bea}{\begin{eqnarray}}
\newcommand{\eea}{\end{eqnarray}}
\newcommand{\beaa}{\begin{eqnarray*}}
\newcommand{\eeaa}{\end{eqnarray*}}
\newcommand{\bmx}{\begin{pmatrix}}
\newcommand{\emx}{\end{pmatrix}}

\newcommand{\zc}{\check z}
\newcommand{\wc}{\check w}

\newcommand{\del}{\partial}

\newcommand{\pd}[1]{{\frac{\del}{\del\smash{ #1}}}}

\newcommand{\half}{\frac{1}{2}}

\newcommand{\nn}{\nonumber}

\newcommand{\8}{{\infty}}

\newcommand{\eps}{\epsilon}

\newcommand{\hc}{{\dagger}}

\newcommand{\ket}[1]{{\,\left|#1\right>}\,}
\newcommand{\kket}[1]{{\, \left|\left|#1\right>\right>\,}}

\newcommand{\cas}{{\mathscr C}}
\newcommand{\am}{{\,\mathrm{am}}}
\newcommand{\dn}{{\,\mathrm{dn}}}
\newcommand{\sn}{{\,\mathrm{sn}}}

\newcommand{\Am}{{\,\mathrm{Am}}}
\newcommand{\Dn}{{\,\mathrm{Dn}}}
\newcommand{\Sn}{{\,\mathrm{Sn}}}
\newcommand{\Cn}{{\,\mathrm{Cn}}}
\newcommand{\id}{{\mathrm{id}}}

\advance\textheight by \topskip
\begin{document}

\baselineskip 17.5pt
\parindent 18pt
\parskip 8pt

\begin{flushright}
\break

DCPT-08/21

\end{flushright}
\vspace{1cm}
\begin{center}
{\LARGE {\bf Covariant particle exchange for }}

{\LARGE {\bf $\kappa$-deformed theories in 1+1 dimensions}}\\[4mm]
\vspace{1.5cm}
{\large  C. A. S. Young\footnote{\texttt{charlesyoung@cantab.net}} and R. Zegers\footnote{\texttt{robin.zegers@durham.ac.uk}}}
\\
\vspace{10mm}

{ \emph{Department of Mathematical Sciences\\ University of Durham\\
South Road, Durham DH1 3LE, UK}}

\end{center}

\vskip 1in
\centerline{\small\bf ABSTRACT}
\centerline{ 
\parbox[t]{5in}{\small 
We consider the exchange of identical scalar particles in theories with $\kappa$-deformed Poincar\'e symmetry. We argue that, at least in 1+1 dimensions, the symmetric group $S_N$ can be realized on the space of $N$-particle states in a $\kappa$-covariant fashion. For the case of two particles this realization is unique: we show that there is only one non-trivial intertwiner, which automatically squares to the identity.}}


\vspace{1cm}

\newpage
\section{Introduction}
The universal enveloping algebra $\mathcal U(P)$ of the Poincar\'e algebra is known to possess a family of deformations
\be\mathcal U_\kappa(P), \qquad\text{with}\qquad \mathcal U_\8(P) = \mathcal U(P) \ee
parameterized by a mass scale $\kappa$ \cite{kPoin,CGST,bicross}. From the dual point of view, in the Hopf-algebraic sense \cite{Majid}, there is a $\kappa$-deformation of the algebra of functions on the Poincar\'e group \cite{SZak}.

Given the existence of this deformation, one would like to know how much of relativistic physics carries over to the $\kappa$-deformed case; in particular, there has been much interest in understanding $\kappa$-deformed quantum field theory \cite{refs,AMFock,GGHMMSS,Rimetal,KMS}. 
Beyond the intrinsic mathematical appeal\footnote{Part of the challenge of $\kappa$-deformation is that, in the language of \cite{Lukierski:2006fv}, $\mathcal U_\kappa(P)$ is thought to be a ``hard'' deformation, as opposed to a twist deformation, of $\mathcal U(P)$ \cite{Lukierski:2006fv,Kosinski:2003xx,LLM}. The algebra of coordinates on $\kappa$-Minkowski space,
\be \left[ x_i, x_j \right] = 0 ,\qquad \left[ x_0, x_i\right] = \frac{1}{\kappa} x_i ,\label{kmink}\ee
is a twist of the commutative case, as in e.g. \cite{BKLVY,GGHMMSS}, but (\ref{kmink}) captures only part of the structure of $\mathcal U_\kappa(P)$.}, a physical motivation for this interest is that $\kappa$-Poincar\'e is known to be a symmetry of the kinematics (i.e. the properties of single particles or free fields -- notably the dispersion relation) of a number of theories, in diverse contexts  \cite{DSR,phonons,AdSCFT}.
It is thus natural to hope that $\kappa$-Poincar\'e is in fact a full dynamical symmetry in certain cases.

In seeking to formulate quantum field theories with $\kappa$-Poincar\'e symmetry, one of the tasks that must be addressed is that of finding the correct modification of the usual algebra
\be \big[ a^\hc(p), a^\hc(q) \big] =0,\quad  \big[a^{}(p),a^{}(q)\big]=0, \quad 
  \big[ a(p) , a^\hc(q) \big] \sim \delta(p-q) \label{aa}\ee
of creation/annihilation operators of field modes. These statements encode the meaning of identical particles: the equation $a^\hc(p)a^\hc(q) =a^\hc(q)a^\hc(p)$ specifies, in particular, that the two tensor product states
\be \ket p \otimes \ket q \quad \text{and} \quad \ket q \otimes \ket p \ee
are to be identified; both describe the same physical state of two identical bosons.
This identification is covariant -- it commutes with the action of the symmetry algebra -- because the coproduct of Poincar\'e is just the usual cocommutative Leibnitz rule
\be \Delta X = 1 \otimes X + X \otimes 1.\ee 
The coproduct of $\kappa$-Poincar\'e, by contrast, is not cocommutative, and it is this which leads to the difficulty in defining states of many identical particles.

This problem was addressed by Daszkiewicz, Lukierski and Woronowicz in \cite{DLW1, DLW2}, at least as far as the non-cocommutativity of the translation generators is concerned. They gave a modified version of the brackets (\ref{aa}), constructed in such a way that the states which are identified have the same total momentum. It is then natural to ask whether there exists a notion of particle exchange which commutes with the entire $\kappa$-Poincar\'e algebra, including the Lorentz sector, and so is frame-independent. 

We gave a partial answer in \cite{IT1}, in which we showed that for the case of two scalar particles, and perturbatively to third order in $\frac{1}{\kappa}$, there is a unique covariant  map exchanging the particles. Of course, one would like to understand why this should be true exactly, rather than just in perturbation theory, and further to know whether particle exchange can still be defined covariantly for states of more than two particles. 
That is, if \be V_m = \{\text{states of a single scalar particle of mass $m$}\}\ee as defined in section \ref{singlep} below, the question is whether there exists for every $N$ a realization of the symmetric group $S_N$ on the $N$-fold tensor product
\be \underset{\text{$N$ times}}{\underbrace{V_m \otimes \dots \otimes V_m}}\ee
with the property that the action of $S_N$ commutes with that of the $\kappa$-Poincar\'e algebra, and reduces as $\kappa\rightarrow \8$ to permutation of the tensor factors so that the standard notion of particle exchange is recovered in the undeformed limit. The space of states of $N$ bosons of this species is then $V_m^{\otimes N}/S_N$. 

In the present work, limiting ourselves to 1+1 dimensions, we will argue that such a definition of particle exchange is possible. We will prove it only for four particles or fewer, but the pattern will by then have become reasonably clear. In the case of two particles, we will find that there is only one non-trivial intertwiner $\tau:V_m\otimes V_m\rightarrow V_m\otimes V_m$.  
For three or more particles, we will show merely that realizations of $S_N$ \emph{exist}; further work is needed to show that one in particular is (we anticipate) picked out in some natural fashion.  

It is worth commenting on the significance of the 1+1 dimensional case. In 1+1 dimensions the concept of identical particles of course becomes rather subtle, because, intuitively speaking, it is not possible to exchange two wave-packets without bringing them close together and letting them interact. This fact underlies much of the rich algebraic structure of integrable quantum field theory (notably Zamolodchikov-Faddeev algebras \cite{ZF}). From a purely 1+1 dimensional perspective, then, it is not obvious that defining a realization of $S_N$ in the sense above is necessary.
Even in 2+1 dimensions exotic statistics are possible, because one can distinguish ``over-crossings'' from ``under-crossings'' when exchanging particles. (See \cite{Schroers} for a discussion of the role of $\kappa$-Poincar\'e in 2+1 dimensional theories of gravity).  
We should therefore stress that here we are treating 1+1 dimensions as a warm-up for 3+1 and higher dimensions: if $S_N$ is to act covariantly in 3+1 dimensions, it is necessary (though not sufficient) that it can do so in 1+1. 

On the subject of exotic statistics, it should also be emphasised that the (unique) intertwiner  $\tau:V_m\otimes V_m\rightarrow V_m\otimes V_m$ we find below \emph{automatically} has the property that $\tau^2=\id$: this is not an extra condition one can choose to place on $\tau$, but is simply true. With quantum Frobenius-Schur duality in mind (see e.g. \cite{ChariPressley}, chapter 10) one might perhaps have expected the intertwiners of representations of $\mathcal U_\kappa(P)$ to obey some sort of Hecke-like relations -- but, luckily for the project of $\kappa$-deformed field theory, this intuition from the case of simple $q$-deformed algebras fails.

The structure of this paper is as follows: after recalling in section \ref{kPdef} the definition of the $1+1$ dimensional $\kappa$-Poincar\'e algebra and the relevance of elliptic rapidity variables, we consider the two-particle case in section \ref{twoparticles}. Then in section \ref{manyparticles} we generalize to states of 3 and 4 particles, before concluding with comments on the general many-particle case and some further discussion in section \ref{conc}. In an appendix we give an extension of the 3+1 dimensional work in \cite{IT1} to the case of three particles.

\section{1+1 dimensional $\kappa$-Poincar\'e, and elliptic rapidities}
\subsubsection*{The Algebra $\mathcal U_\kappa(P)$}\label{kPdef}
For the present purposes it is most convenient to work in the original basis of \cite{kPoin,CGST}, rather than the bicrossproduct basis subsequently found in \cite{bicross}. In the original basis the two-dimensional $\kappa$-Poincar\'e algebra $\mathcal U_\kappa(P)$ is defined by 
\be \left[ N, P_0 \right] = P_1, \qquad \left[ N, P_1 \right] = \kappa \sinh \frac{P_0}{\kappa} \qquad \left[ P_0, P_1 \right] = 0,\label{kP}\ee
where $N$ is the generator of boosts and $P_\mu$, $\mu=1,2$, are the momentum generators. 
The coalgebra is given by
\bea \Delta N &=& e^{-\frac{P_0}{2\kappa}} \otimes N + N \otimes e^{\frac{P_0}{2\kappa}} \label{coalg1}\\
 \Delta P_1 &=& e^{-\frac{P_0}{2\kappa}} \otimes P_1 + P_1 \otimes e^{\frac{P_0}{2\kappa}} \\
 \Delta P_0 &=& 1 \otimes P_0 + P_0 \otimes 1 \label{coalg3}\eea
and, for completeness, the antipode and counit maps are 
\be S N = -N+\frac{1}{2\kappa} P_1,\qquad SP_0 = -P_0,\qquad SP_1 = -P_1 ,\ee
\be \eps N = 0, \qquad \eps P_0 = 0, \qquad \eps P_1 = 0.\ee
The Casimir takes the form
\be \cas = \left(2\kappa \sinh \frac{P_0}{2\kappa} \right)^2 - P_1^2.\label{cas}\ee
Although much of what follows will apply for general $\kappa\in\mathbb C$, where it is relevant -- e.g. in the shape of the contours of $\cas$ in section \ref{manyparticles} -- we should stress that we consider only the case 
\be \kappa \in \mathbb R, \qquad \kappa>1. \ee

\subsubsection*{Single particles}\label{singlep}
The basis above lends itself to a natural parameterization of single-particle states in terms of Jacobi elliptic functions. Let $V_m$ be the space of states of a single scalar particle of mass $m$, $m^2>0$, spanned by the momentum modes
\be \ket{p_0,p_1}, \quad \text{with}\quad P_\mu \ket{p_0,p_1} = p_\mu \ket{p_0,p_1},\quad \cas\ket{p_0,p_1} = m^2\ket{p_0,p_1} . \label{modes}\ee
Then on wavefunctions $\psi(p_0,p_1)$ of states  
\be \ket\psi = \int d^2p\, \delta\left(m^2-4\kappa^2 \sinh^2 \frac{p_0}{2\kappa} + p_1^2\right)\ket{p_0,p_1} \psi(p_0,p_1)\,\,\in V_m \ee
the algebra generators are realized as\footnote{Here and in what follows, statements like ``$P_0=p_0$'' are of course really shorthand for $\rho(P_0) \psi(p)= p_0 \psi(p)$ where $P_0 \ket\psi =: \int d^2p\,\delta\left(m^2-4\kappa^2 \sinh^2 \frac{p_0}{2\kappa} + p_1^2\right)\ket{p_0,p_1} \rho(P_0) \psi.$}
\be P_0 = p_0,\qquad P_1 = p_1, 
\qquad N  = \kappa \sinh \left(\frac{p_0}{\kappa}\right) \pd{p_1} + p_1 \pd{p_0}\ee
-- the final equation following from the defining relations (\ref{kP}). 
The basis modes can be parameterized by a single ``elliptic rapidity'' or ``uniformizing'' variable $z$, \cite{rapidity} 
\be \ket{z} \equiv \ket{p_0(z),p_1(z)},\ee 
defined by the demand that the boost generator be realized as
\be N = \pd z \ .\label{unif}\ee
This yields differential equations for $p_0(z)$, $p_1(z)$, which are solved by setting
\be p_0(z) = -km \am\left(\frac{iz}{k} + iK'(k) - K(k) ; k\right) \label{p0z}\ee
\be p_1(z) = - im \dn\left(\frac{iz}{k} + i K'(k) - K(k); k\right) \label{p1z}\ee
where the modulus is 
\be k=2i\kappa/m\ee and $K(k)$, $iK'(k)$ are the elliptic quarter-periods. For an overview of the theory of elliptic functions see for example \cite{HTF}.
The shifts in the arguments here are chosen such that $p_\mu(z)$ is real for real $z$ and agrees in the limit $\kappa\rightarrow\8$ with the usual parameterization
\be p_0(z) \underset{\kappa\rightarrow \8}\sim m \cosh z \ee
\be p_1(z)  \underset{\kappa\rightarrow \8}\sim m\sinh z \ee
of the mass shell.
For finite $\kappa$ we restrict the functions $p_\mu(z)$ to the finite domain (note that $k\in i\mathbb R\Rightarrow K\in \mathbb R$)
\be ikK < z < -ikK \ee 
at the end-points of which they have poles.

In what follows it will be useful to use the shorthand notation
\be \Sn(z) := \sn\left(\frac{iz}{k} + i K'(k) - K(k); k\right) \qquad ( = -\frac{1}{k} \mathrm{dc}\left(\frac{iz}{k},k\right)) \label{Sndef}\ee
and likewise for $\Cn$ and $\Dn$. 
 
\subsubsection*{Tensor products}
The tensor product $V_m \otimes V_{m}$
carries, as usual, a representation of $\mathcal U_\kappa(P)$ specified by the coalgebra (\ref{coalg1}-\ref{coalg3}). On the wavefunctions $\psi(r_0,r_1,s_0,s_1)$ of states
\bea \ket\psi &=& \int d^2r d^2s\,\delta(m^2-4\kappa^2 \sinh^2 \frac{r_0}{2\kappa} + r_1^2)\delta(m^2-4\kappa^2 \sinh^2 \frac{s_0}{2\kappa} + s_1^2)\nn\\&&\qquad\qquad\qquad\ket{r_0,r_1}\otimes \ket{s_0,s_1} \psi(r_0,r_1,s_0,s_1)  \eea
in $V_m\otimes V_{m}$ the generators are realized as
\be P_0=  r_0 + s_0, \qquad P_1 = e^{-\frac{r_0}{2\kappa}}s_1 +  r_1e^{\frac{s_0}{2\kappa}}\label{deltaP}\ee
and 
\be N = e^{-\frac{r_0}{2\kappa}}\pd w  + e^{\frac{s_0}{2\kappa}}\pd z\label{2boost}\ee
where $z,w$ are the elliptic rapidities of the momenta $r_\mu = r_\mu(z)$ and $s_\mu=s_\mu(w)$.
This extends in an obvious fashion to give the action of the $\kappa$-Poincar\'e algebra $\mathcal U_\kappa(P)$ on the tensor product of three or more single-particle representations.
\section{States of two identical particles}\label{twoparticles}
Our concern here is with the covariant definition of states of many \emph{identical} particles, of mass say $m$. As discussed in the introduction this amounts to showing that there is a suitable realization of the symmetric group $S_N$ on the $N$-fold tensor product $V_m^{\otimes N}$.
We begin with the case of two particles, and prove the following
\begin{claim}\label{2pclaim} For any mass $m$ there is a unique map
\be \tau : V_m \otimes V_{m} \longrightarrow V_{m} \otimes V_m \ee
commuting with the action of $\mathcal U_\kappa(P)$ and such that $\tau\neq \id$. It obeys
\be \tau^2 = \id\label{tau2=1}.\ee\end{claim}
In \cite{IT1} we showed this in $3+1$ dimensions to the first few orders in $\frac{1}{\kappa}$ by writing $\tau : \ket{r_\mu}\otimes\ket{s_\mu} \rightarrow \ket{f_\mu(r,s)}\otimes\ket{g_\mu(r,s)}$ and solving for the functions $f$ and $g$ explicitly.\footnote{We failed to observe in \cite{IT1} that $\tau^2=\id$ holds automatically when the other conditions listed in that paper are imposed, but have now checked that this is so.} This has the merit of concreteness, but it obscures the symmetry implicit in $\tau^2=\id$. We want to show here that, in the 1+1-dimensional case at least, the claim is true exactly rather than just perturbatively, so we approach the problem from a point of view in which this symmetry is manifest from the start. 

The tensor product $V_m\otimes V_m$ decomposes into a direct sum of representations $U_M$,
\be V_m \otimes V_{m} = \int^{\8}_{M_0} U_MdM ,\ee
labelled by the total value of the Casimir:
\be (\Delta \cas) \ket z \otimes \ket w = M^2 \ket z \otimes \ket w\quad \Leftrightarrow\quad\ket z \otimes \ket w\in U_M.\ee 
The lower bound in the integral is $M_0= 2\kappa\sinh \left(2\sinh^{-1}\frac{m}{2\kappa}\right)$. 
If we can show that for every $M>M_0$ the representation $U_M$ consists of the direct sum 
\be U_M \cong V^{(1)}_M \oplus V^{(2)}_M \label{split}\ee
of precisely two copies of the irrep $V_M$, while at the lower bound 
\be U_{M_0}\cong V_{M_0},\ee
then we are effectively done, because the map we seek is necessarily the map exchanging the copies for all $M>M_0$ and acting trivially on the irreducible component $U_{M_0}$.\footnote{Strictly, to ensure $\tau$ is non-trivial, it suffices that it exchange the copies for some $M$ and we should appeal to continuity requirements -- which we have not made explicit in claim \ref{2pclaim} -- to show that it does so for all values of $M>M_0$.}

This becomes very familiar in the undeformed limit $\kappa=\8\Leftrightarrow k=\8$, in which $M_0$ is the usual mass threshold $2m$. The $(z,w)$ plane can be sketched as follows:
\begin{center} \epsfig{file=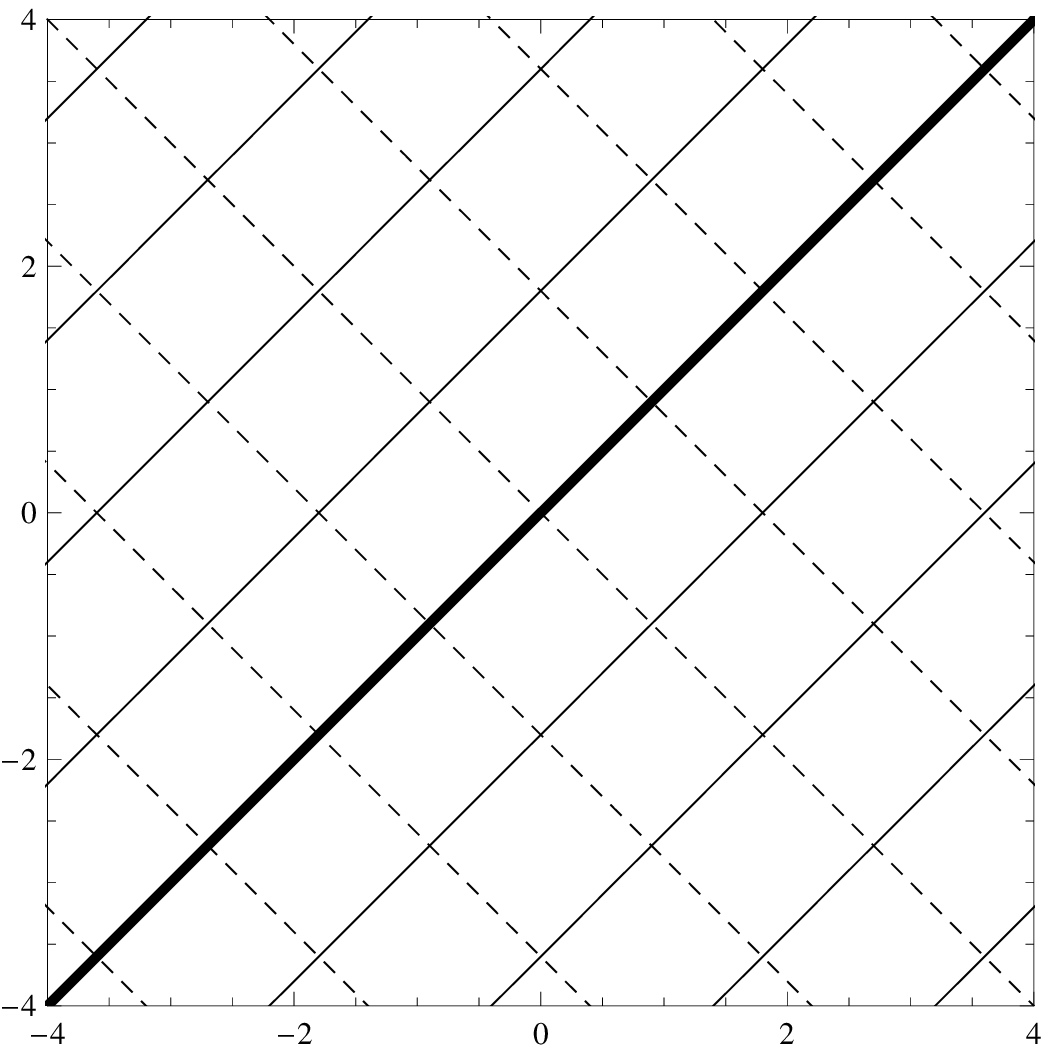,height=3in} \end{center}
The boost generator is the constant vector field $\del_z+\del_w$, finite boosts are rigid translations in this direction, and the (bases of the) irreducible components of $V_m\otimes V_m$ are the solid diagonal lines labelled by the value of the rapidity difference
\be \mu_{\kappa=\8} := \frac{z-w}{2}.\ee
Exchanging the particles, in the undeformed sense $\tau_{\kappa=\8}:z\leftrightarrow w$, flips the sign of $\mu$ while preserving $M= 2m\cosh\half(z-w)$. We have drawn the contour $\mu=0$ in bold, and the contours of $\half(z+w)$ as dashed lines. 

In the same spirit, when $\kappa<\8$ the statements above about the $U_M$ are true if there exists a change of coordinates \be (z,w) \mapsto (\mu(z,w), \phi(z,w))\label{chart}\ee on the space of basis states $\ket z \otimes \ket w$ such that the value $M^2$ of the Casimir is independent of $\phi$ and an even, convex function of $\mu$.
The coordinate $\mu$ should be invariant under boosts:
\be  0 =N\mu = \left(e^{-\frac{r_0}{2\kappa}}\pd w  + e^{\frac{s_0}{2\kappa}}\pd z\right)\mu.\label{mueqn}\ee
That is, in the notation introduced above in (\ref{Sndef}),
\be \left[ \left( \Cn z + i\Sn z \right)\pd w + \left( \Cn w  - i \Sn w\right) \pd z\right] \mu = 0 \ee
which may be solved by separation of variables to find
\bea \mu &=& -\half \log \frac{1}{k}\left( \Dn z - k\Cn z\right)\left( \Dn z - i k \Sn z\right)\\
         && -\half \log \frac{1}{k}\left(  \Dn w - k\Cn w\right)\left( \Dn w + i k \Sn w\right)
          +\half \log \left( 1-\frac{1}{k^2} \right) \nn.\eea
Here the overall factor is chosen so that $\mu\rightarrow \half(z-w)$ as $\kappa\rightarrow \8$, and the final term, independent of $z,w$ and vanishing in the undeformed limit, is included to give the correct origin for $\mu$ in what follows. On rearranging and noting the standard identities for Jacobi elliptic functions $$\left(\Dn z - ik \Sn z\right) \left(\Dn z + ik \Sn z\right)=1 \quad\text{and}\quad \left(\Dn z - k \Cn z\right) \left(\Dn z + k \Cn z\right)=1-k^2,$$ one has that
\be \left(1-\frac{1}{k^2}\right) e^{-2\mu}  
  = \frac{1}{k^2} \left(\Dn z - k \Cn z\right) \left(\Dn z - ik \Sn z\right)
                    \left(\Dn w - k \Cn w\right) \left(\Dn w + ik \Sn w\right) \ee
\be \left(1-\frac{1}{k^2}\right) e^{2\mu} =\frac{1}{k^2} \left(\Dn z + k \Cn z\right) \left(\Dn z + ik \Sn z\right)
                    \left(\Dn w + k \Cn w\right) \left(\Dn w - ik \Sn w\right) .\ee  
Then, after some further use of standard identities in the second step,
\bea \left(1-\frac{1}{k^2}\right) \left(e^{2\mu}+e^{-2\mu}\right)
  &=& 2 \Dn z\Dn w \left( \Cn z \Cn w + \Sn z \Sn w - i \Sn w \Cn z + i \Sn z \Cn w \right) \nn\\
&&+ 2i \Dn^2 w \Sn z \Cn z - 2i \Dn^2 z\Sn w \Cn w\nn \\&&+ \frac{2}{k^2} \Dn^2 w \Dn^2 z + 2k^2 \Sn z \Cn z \Sn w \Cn w \\
&=& \left( \left(\Cn z+ i \Sn \right) \Dn w + \left( \Cn w - i \Sn w\right) \Dn z \right)^2 \nn\\&& + k^2 \left( \Sn z \Cn w + \Cn z \Sn w\right)^2 - 2\left( 1- \frac{1}{k^2}\right), \label{Cder}\eea 
and here, given (\ref{cas}), (\ref{p0z}-\ref{p1z}) and (\ref{deltaP}), one recognizes the Casimir appearing on the right-hand side, yielding finally
\be \left(1-\frac{1}{k^2} \right) \left(2\cosh \mu\right)^2 = \frac{M^2}{m^2} \ee
\be \Rightarrow M = 2m\cosh \mu \, \sqrt{1-\frac{1}{k^2} },\label{C12}\ee
which agrees with the standard $\kappa=\8$ relation up to the $k\sim\kappa$-dependent factor. 
This coordinate $\mu(z,w)$ is by definition unique up to reparametrization $\mu\mapsto\tilde\mu(\mu)$ by an odd, strictly monotonic, function $\tilde\mu$. 
To specify the change of coordinates (\ref{chart}), and hence the map $\tau$, it remains to choose a suitable function $\phi(z,w)$. Here there is more freedom, and (at least) three possible choices are worth commenting on. 

For \emph{any} $\phi(z,w)$ such that $(\mu,\phi)$ is a good chart, the map 
\be \tau: (\mu,\phi) \leftrightarrow (-\mu, \phi). \label{taudef}\ee 
defines, for each $\mu>0$, a bijection between the pair of contours $\pm\mu$ and hence between the bases of the two copies of $V_{M(\mu)}$ in (\ref{split}). But, to commute with the action of $\mathcal U_\kappa(P)$, $\tau$ must identify the states correctly, sending the state with a given total momentum in one copy to the state with the same total momentum in the other. The simplest way to ensure this is simply to choose $\phi$ to be $P_1(z,w)$. 
This possibility is illustrated below, with $k=25i$. As above, contours of $\mu(z,w)$, and hence of the Casimir, are drawn as continuous lines, with $\mu=0$ in bold. Contours of $P_1$ (left) and $P_0$ (right) are drawn as dashed lines. (For clarity of the plots, equally spaced contours of $\sinh^{-1}(P_i)$ are shown.)  
\begin{center}\epsfig{file=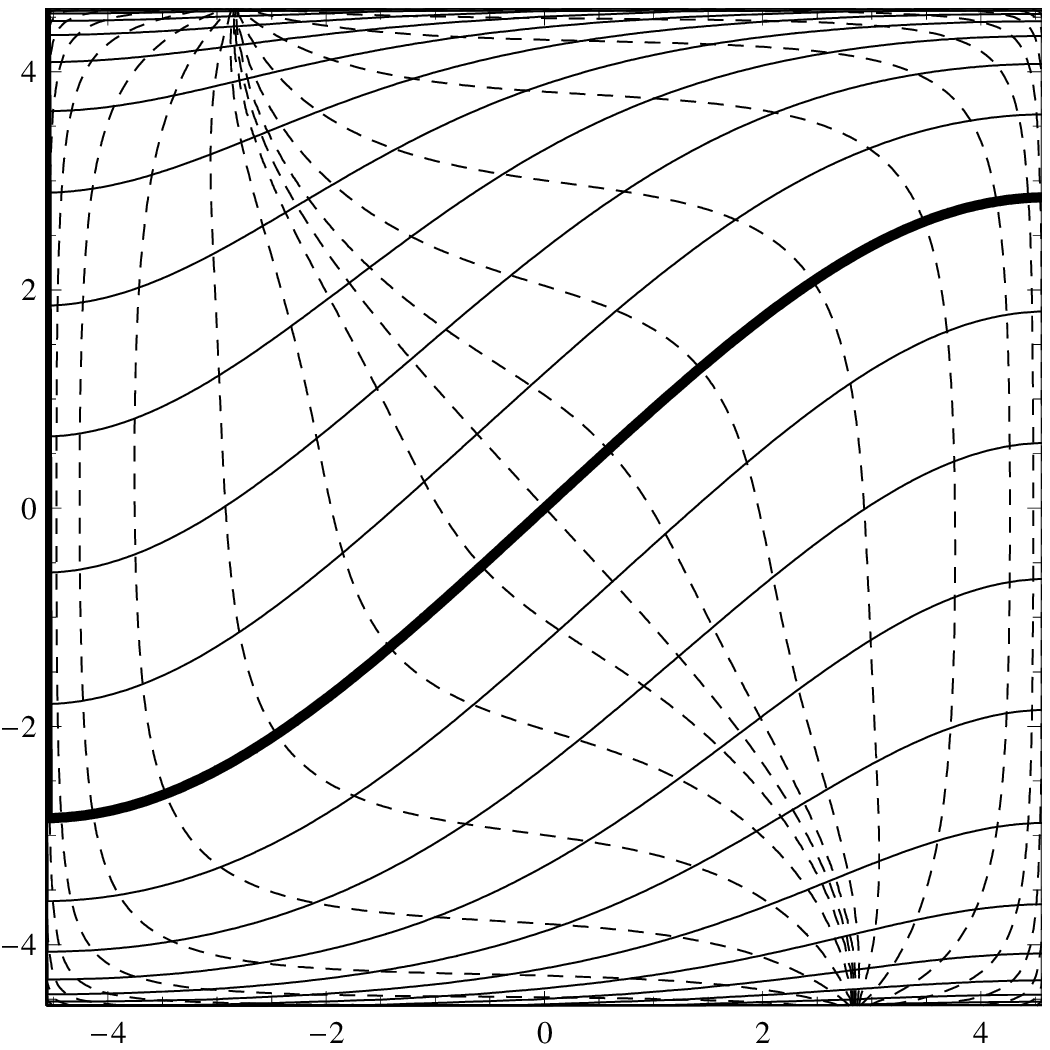,height=3in}$\quad$ \epsfig{file=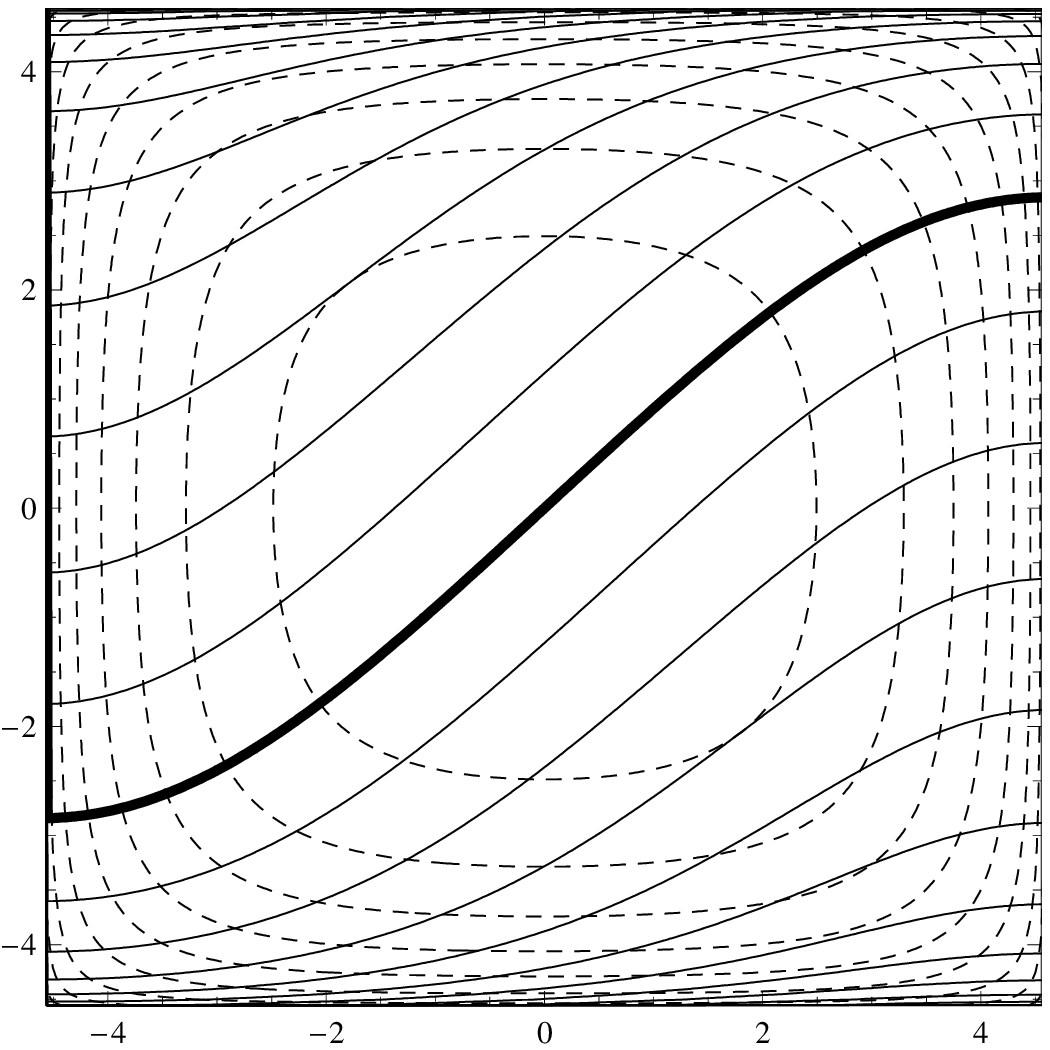,height = 3in}\end{center}
A second choice of chart on the physical region
\be ikK < z < -ikK, \qquad ikK < w< -ikK \label{physreg}\ee
of parameter space is as follows. Given the equation (\ref{mueqn}) for $\mu$, it is natural to consider in addition to
\be N = e^{-\frac{r_0}{2\kappa}}\pd w  + e^{\frac{s_0}{2\kappa}}\pd z
      = \left( \Cn z + i\Sn z \right)\pd w + \left( \Cn w  - i \Sn w\right) \pd z\ee
the vector field
\be \tilde N = e^{\frac{r_0}{2\kappa}}\pd w  - e^{-\frac{s_0}{2\kappa}}\pd z
      = \left( \Cn z - i\Sn z \right)\pd w - \left( \Cn w  + i \Sn w\right) \pd z.\ee
By construction $N$ and $\tilde N$ define a smoothly-varying orthogonal (with respect to the metric $dz^2 + dw^2$) frame  everywhere in the region (\ref{physreg}). It follows that the functions $\mu$ and $\phi$ whose contours are integral curves of $N$ and $\tilde N$ respectively are good coordinates. 
A suitable solution to $\tilde N\phi=0$ is
\bea \phi &=& -\half\log\frac{1}{k} 
\left( \Dn z - k\Cn z\right)\left( \Dn z + ik\Sn z\right) \nn\\
&& + \half\log\frac{1}{k} \left( \Dn w - k\Cn w\right)\left( \Dn w - ik\Sn w \right) +\frac{i\pi}{2}.\eea
One may then compute
\bea 2\kappa \sinh \frac{P_0}{2\kappa} &=&  2m\cosh \mu \cosh \phi \,\sqrt{1-\frac{1}{k^2}} \nn \\
P_1 &=& 2m\cosh \mu \sinh \phi \,\sqrt{1-\frac{1}{k^2}} \label{sce}\eea
and so it emerges that the total momentum $P_\mu$ is invariant under (\ref{taudef}) for this choice of $\phi$. With these coordinates, $(\mu,\phi)$, we have therefore proven claim \ref{2pclaim} above. 

The contours of $\mu$ and $\phi$ have the following shape (once more with $k=25i$), which illustrates clearly how the action of $\tau$ is ``warped'' by the $\kappa$-deformation:
\begin{center}\epsfig{file=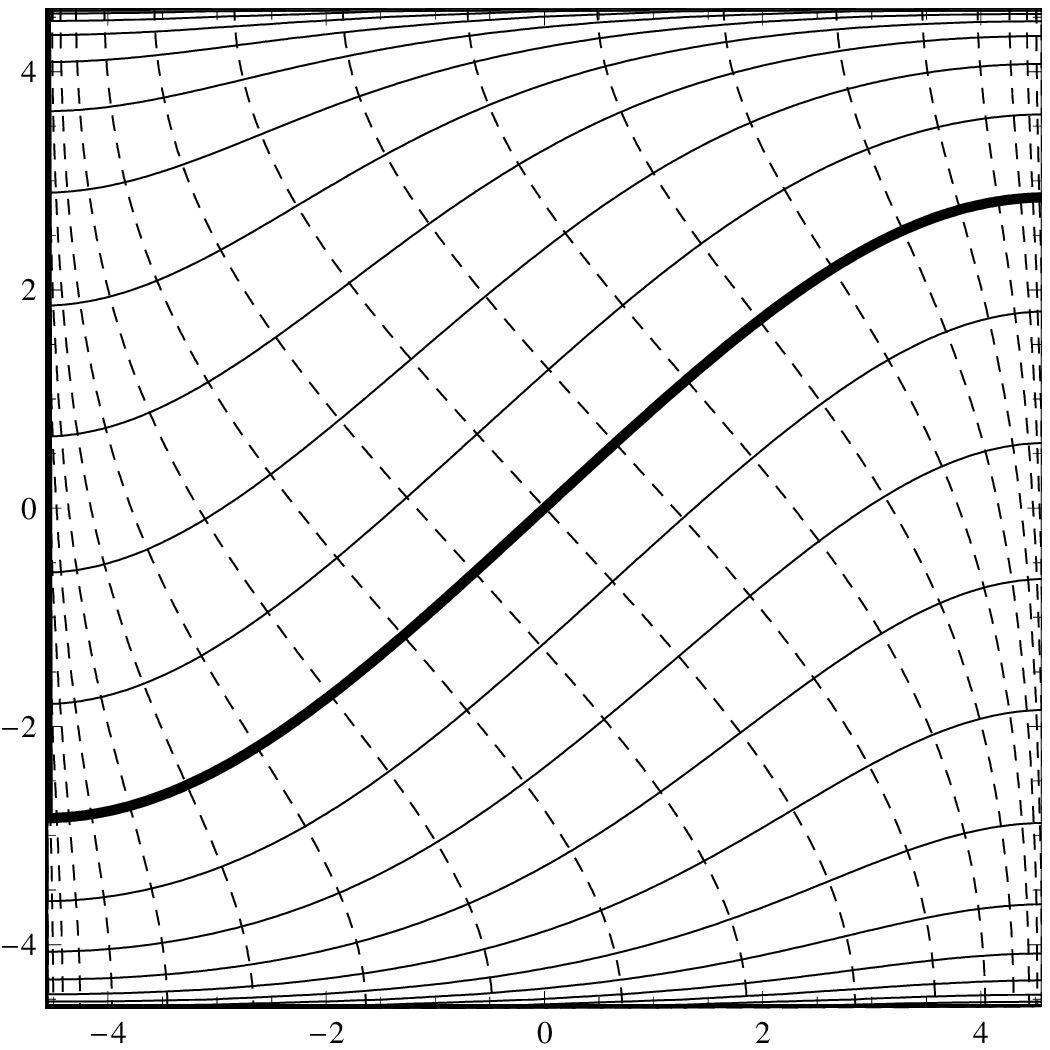,height=3in}\end{center}

The similarity of (\ref{sce}) and (\ref{C12}) to the usual undeformed expressions suggests that the following quantities are in a sense natural ``two-particle-deformed'' corrections to the rapidities $z,w$.   
\be z \underset{\8\leftarrow\kappa} \leftarrow \mu - \phi = -\log\frac{1}{k} \left(\Dn w - k \Cn w\right) \left( \Dn z - ik \Sn z\right)  + \half\log\left(1-\frac{1}{k^2}\right) + \frac{i\pi}{2} \ee
\be w \underset{\8\leftarrow\kappa} \leftarrow \mu + \phi = -\log\frac{1}{k} \left(\Dn z - k \Cn z\right) \left( \Dn w + ik \Sn w\right)  + \half\log\left(1-\frac{1}{k^2}\right) - \frac{i\pi}{2} \ee
On the other hand, it is worth remarking that there is in principle a third choice of  coordinates, $(\mu,\xi)$ say, with respect to which the two-particle boost operator has the form
\be N = \pd \xi .\ee 
(For the $\phi$ above we have only that $N\sim \pd \phi$: to be precise $N= -\frac{2i}{k} \sqrt{1-\frac{1}{k^2}} \cosh \mu \cosh \phi \pd \phi$.) \emph{This} coordinate $\xi$ is the true elliptic rapidity of the two-particle state, in the sense discussed for one-particle states above; but working with it explicitly is rather awkward because one has to manipulate elliptic functions whose modulus is $\mu$-dependent. 

\subsection{Additive momentum labels for two-particle states}
Before moving on from the 2-particle to the many-particle case, let us digress briefly on the question of \emph{momentum labels} for 2-particle states. 
One of our goals throughout is to keep as much of the usual structure from the undeformed case as possible, in order to find out which modifications are genuinely forced on us by the $\kappa$-deformation. 
So far we have established the existence of the intertwiner $\tau$, which ensures that there is a covariant definition  
\be V_m \otimes V_m \Big/ \tau \ee
of the space of two-particle states such that the \emph{counting} of states agrees with the $\kappa=\8$ result.  
In the undeformed case one also has a clear notion of the (unordered pair of) individual momenta 
\be \{ p_1, p_2\} \ee
of the constituent particles of the state
\be \ket{p_1,p_2} \ee 
because, obviously, exchange of particles merely permutes these momentum labels. Here the action of $\tau$ is more complicated, but, following \cite{DLW1,DLW2}, it is natural to ask whether nevertheless there exists a labelling of tensor product states\footnote{Our notation is related to that of \cite{DLW1} by (writing creation operators as $a^\hc$)
$$  \ket r\otimes \ket s = a^\hc (r) a^\hc(s) \ket 0, \qquad \kket{p,q}= a^\hc(p) \circ a^\hc(q) \ket{0}.$$ In that paper eqns (\ref{pqadd}) were solved by setting $a^\hc(p) \circ a^\hc(q) = a^\hc(p_0,e^{-\frac{q_0}{2\kappa}} p_1) a^\hc(q_0,e^{\frac{p_0}{2\kappa}} q_1)$. Imposing the relation $[ a^\hc(p)\, \overset{\circ}{,}\, a^\hc(q) ] = 0$ then identifies pairs of states in $V_m\otimes V_m$ of equal total momentum, although not in a boost-covariant fashion.}
\be \kket{p,q} := \ket{r(p,q)}\otimes \ket{s(p,q)}\in V_m\otimes V_m \label{braided}\ee
in which the two-vectors $r$ and $s$ depend on $p=(p_0,p_1)$ and $q=(q_0,q_1)$ in such a way that 
\be \tau \kket{p,q} = \kket{q,p}\label{pqflip} \ee
and the total momentum is additive:
\bea \Delta P_0 \kket{p,q} = \left(r_0(p,q) + s_0(p,q)\right)\kket{p,q} &=& (p_0 + q_0) \kket{p,q}\nn\\ 
\Delta P_1 \kket{p,q} = \left(r_1(p,q)e^{\frac{s_0(p,q)}{2\kappa}} + e^{-\frac{r_0(p,q)}{2\kappa}}s_1(p,q)\right)\kket{p,q} &=& (p_1 + q_1) \kket{p,q} .\label{pqadd}\eea

Let us consider placing on the labels $p,q$ the further condition that the single-particle states $\ket p$ and $\ket q$ should both be on-shell with mass $m$. This means $p$ and $q$ are functions of rapidities, $\zc$ and $\wc$ say:
\be p_0 = -km\Am \zc \qquad q_0=-km \Am \wc \ee
\be p_1 = -im\Dn \zc \qquad q_1=-im \Dn \wc \ee
as in (\ref{p0z}-\ref{p1z}). Finding the functions $r(p,q)$ and $s(p,q)$ in (\ref{braided}) is then a matter of expressing the rapidities $z$ and $w$ (of $r$ and $s$, as before) in terms of $\zc$ and $\wc$. Now, in view of (\ref{pqadd}) we have 
\bea \Am \zc + \Am \wc &=& \Am z + \Am w \label{aaaa}\\
 \Dn \zc + \Dn \wc &=& \Dn z \left( \Cn w - i \Sn w\right) + \Dn w \left(\Cn z + i \Sn z\right).\label{dndn}\eea
Applying the boost operator $N= \left(\Cn w - i \Sn w\right) \pd z + \left( \Cn z + i \Sn z\right)\pd w $ to these one finds
\be \Dn \zc\, N \zc + \Dn \wc \,N \wc = \Dn z \,N z + \Dn w \,N w = \Dn \zc + \Dn \wc \ee
\be \Sn \zc \Cn \zc \, N \zc + \Sn \wc \Cn \wc \, N \wc 
       =  \Sn z \Cn z(\Cn^2 w - \Sn^2 w)   +\Sn w \Cn w (\Cn^2 z -\Sn^2 z), \ee
after some straightforward manipulations in the second case. On the right here $z$ and $w$ may be replaced throughout by their checked counterparts. (Consider doubling eqn (\ref{aaaa}) and taking the sine.)  
One then has two linear equations for $N\zc$ and $N\wc$ in terms of the checked variables only, and crucially the symmetry of these equations is such that the solutions for $N\zc$ and $N\wc$ are related by flipping $\zc\leftrightarrow \wc$. 
Therefore the boost operator
\be N= N\zc \pd \zc + N\wc \pd \wc \ee
is symmetric under $\zc\leftrightarrow \wc$, or equivalently under $p\leftrightarrow q$. Exchanging $p\leftrightarrow q$ therefore commutes with boosting, so this must indeed be how the (unique) intertwiner $\tau$ of claim \ref{2pclaim} acts in these coordinates. Thus, by demanding that $p$ and $q$ are individually on-shell, we have obtained (\ref{pqflip}) for free.     

At first sight this is very appealing, because it seems to define a labelling of two-particle states in which $\kappa$-deformed particle exchange acts in a simple fashion, and in which momentum is additive. This is indeed true for all those states in $V_m\otimes V_m$ 
which \emph{can} be so labelled, but there is an important caveat: the change of variables $(\zc,\wc)\leftrightarrow (z,w)$ is not everywhere well-defined. This can be seen from the shapes of the contours of the total Casimir (solid lines) and momentum $P_1$ (dashed lines) in the $(\zc,\wc)$ plane, drawn below with $k= 100i$,
\begin{center}\epsfig{file=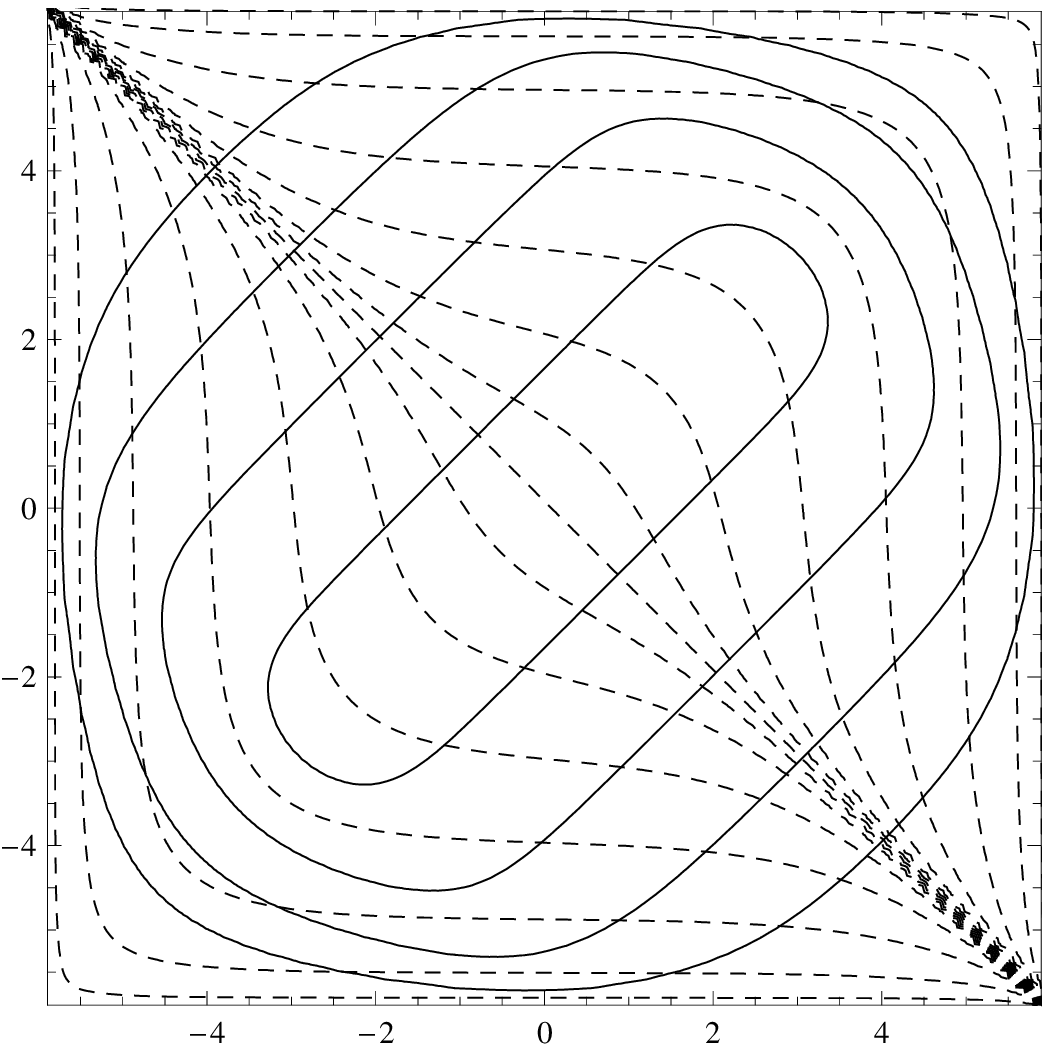,height=3in}\end{center}
Recall that in the $(z,w)$ plane each contour of $M^2=C_{12}$ intersects each contour of $P_1$ exactly twice (and $\tau$ exchanges these intersection points). By contrast, one sees that in the $(\zc,\wc)$ plane the contours of $M$ are closed curves and can intersect contours of $P_1$ any number of times between zero and four. Some states in $V_m \otimes V_m$ therefore correspond to more than one pair $(\zc,\wc)$ -- to two, in generic cases -- and, perhaps more importantly, some states are not represented by any pair $(\zc,\wc)$.

\section{More than 2 particles}\label{manyparticles}
\begin{figure}
$\,\,\,$\epsfig{file=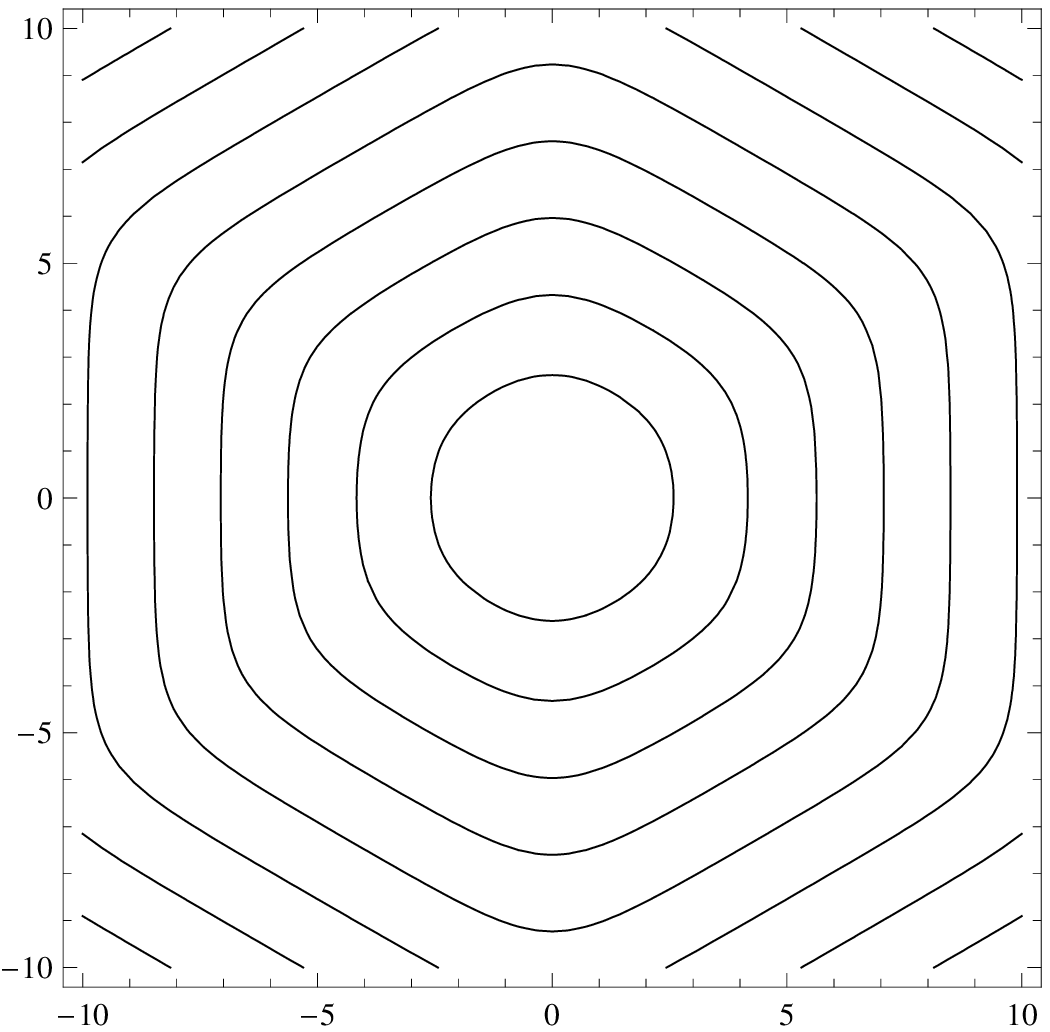,height=3in}$\quad$\epsfig{file=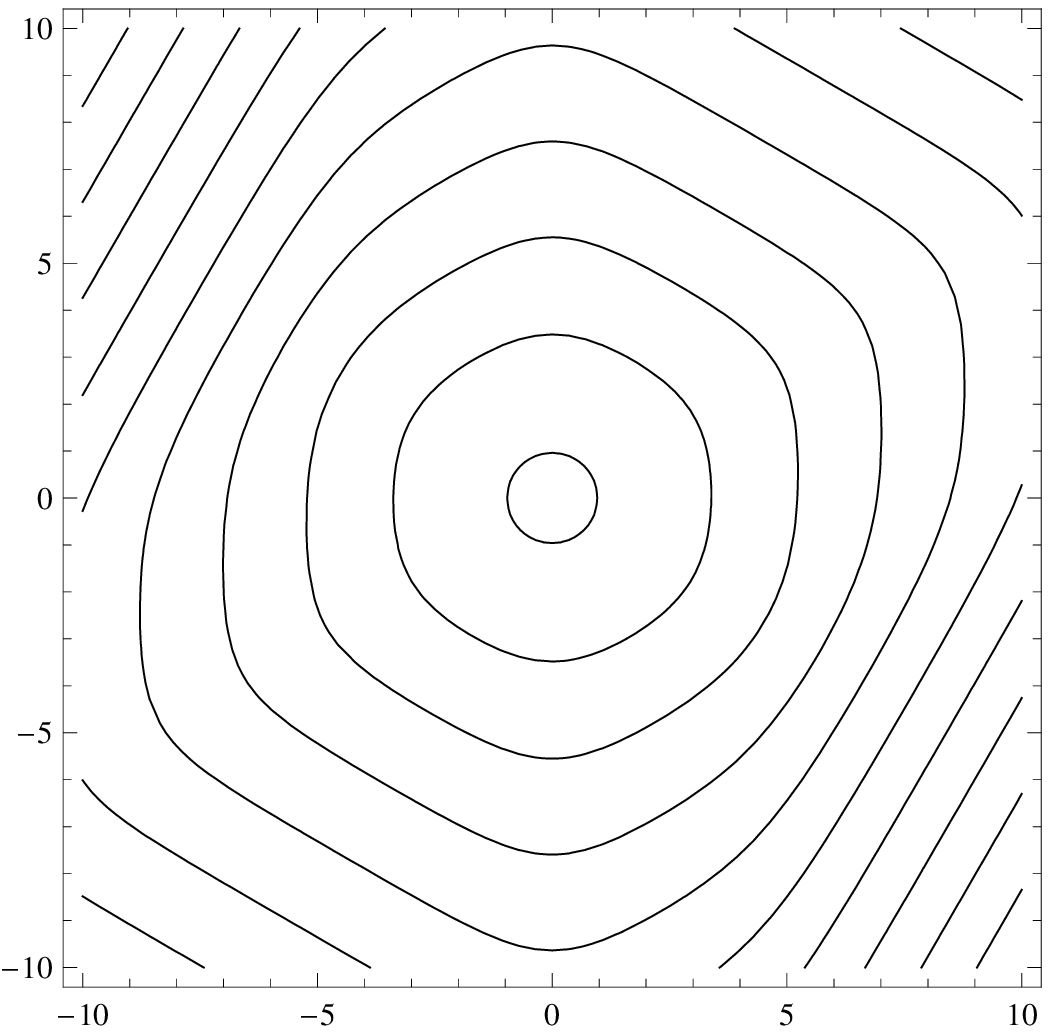,height=3in}

$ $

\epsfig{file=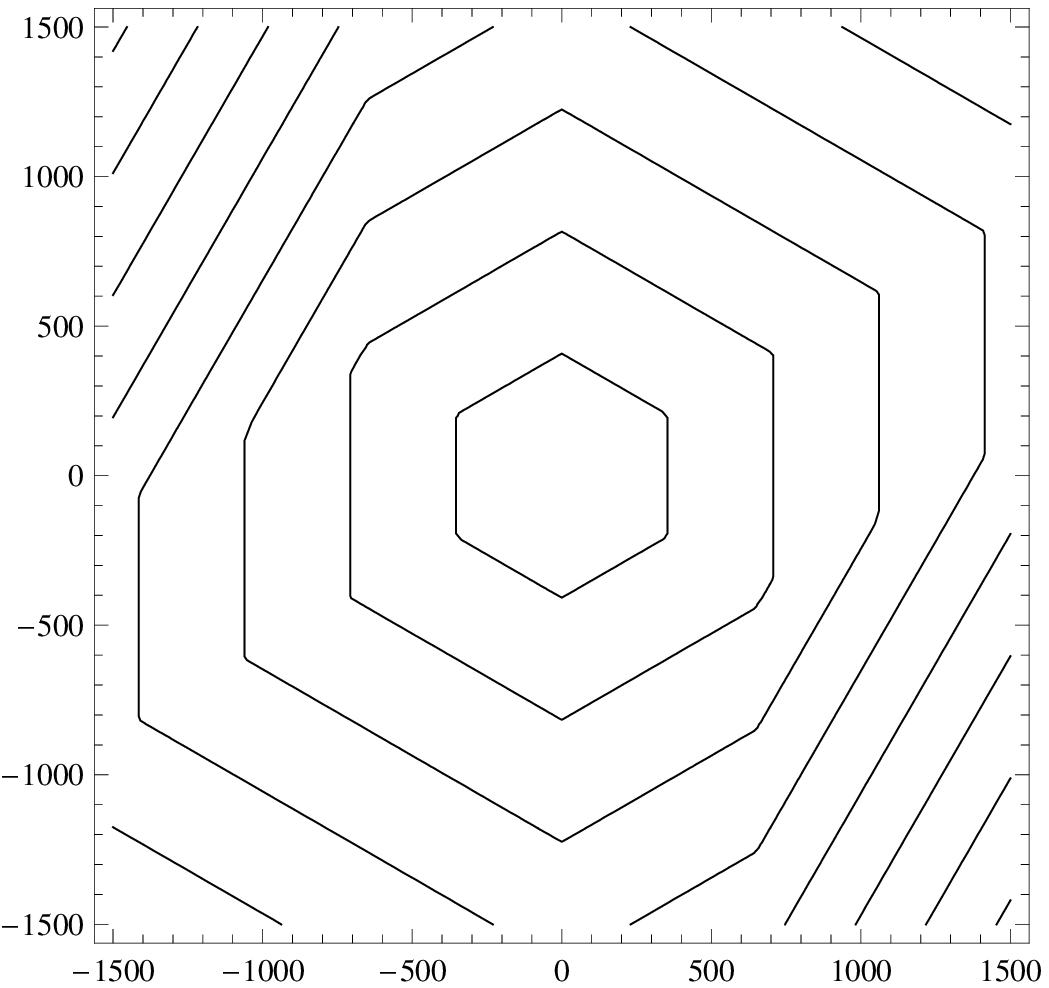,height=3in}$\quad$\epsfig{file=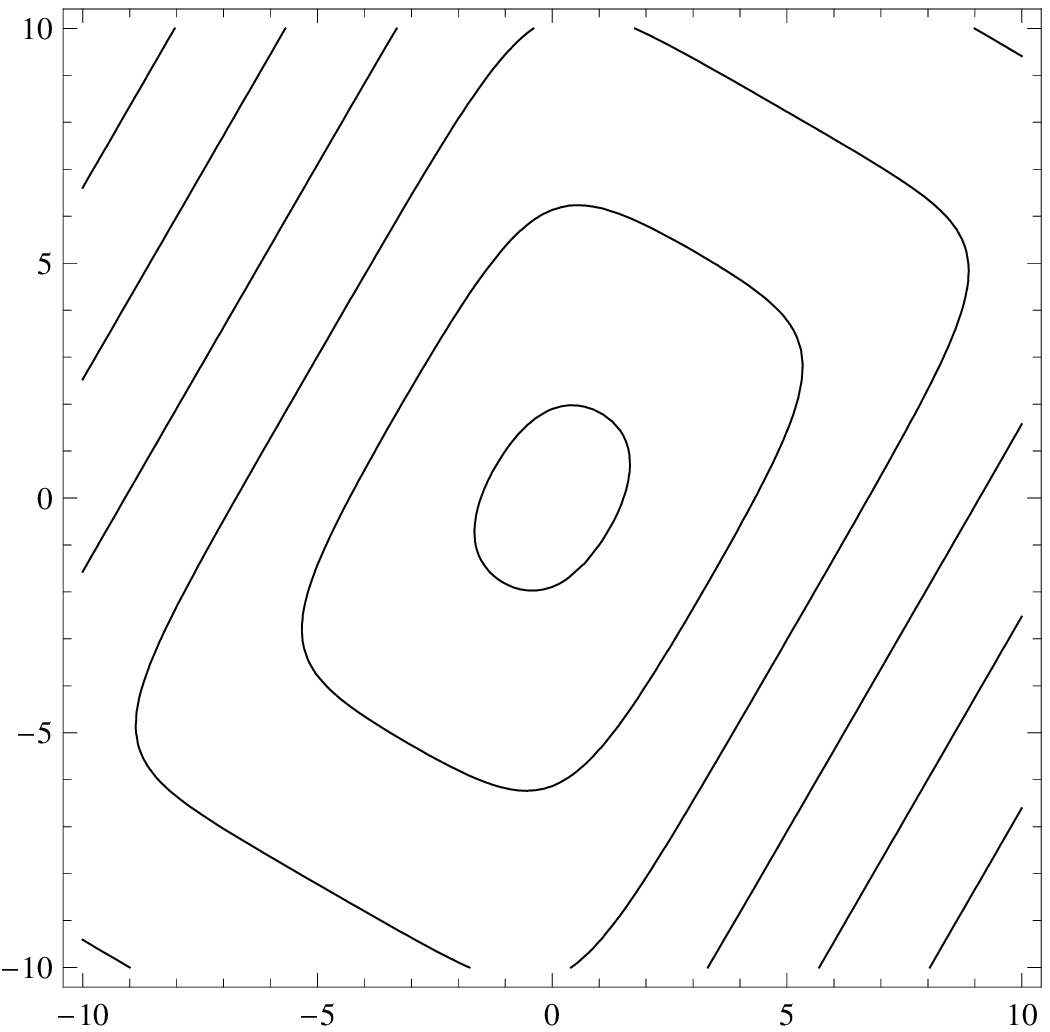,height = 3in}
\caption{Contours of $\log\left(C_{123}/m^2\right)$ with, clockwise from top left, $k=\8$, $25i$, $i$. The lower left plot, drawn to a different scale, has $k=10^{200}i$ and illustrates the approximately piecewise linear shape of the contours when $C_{123}$ is large.}\label{C3pic}
\end{figure}

\begin{figure}
$\,\,\,$\epsfig{file=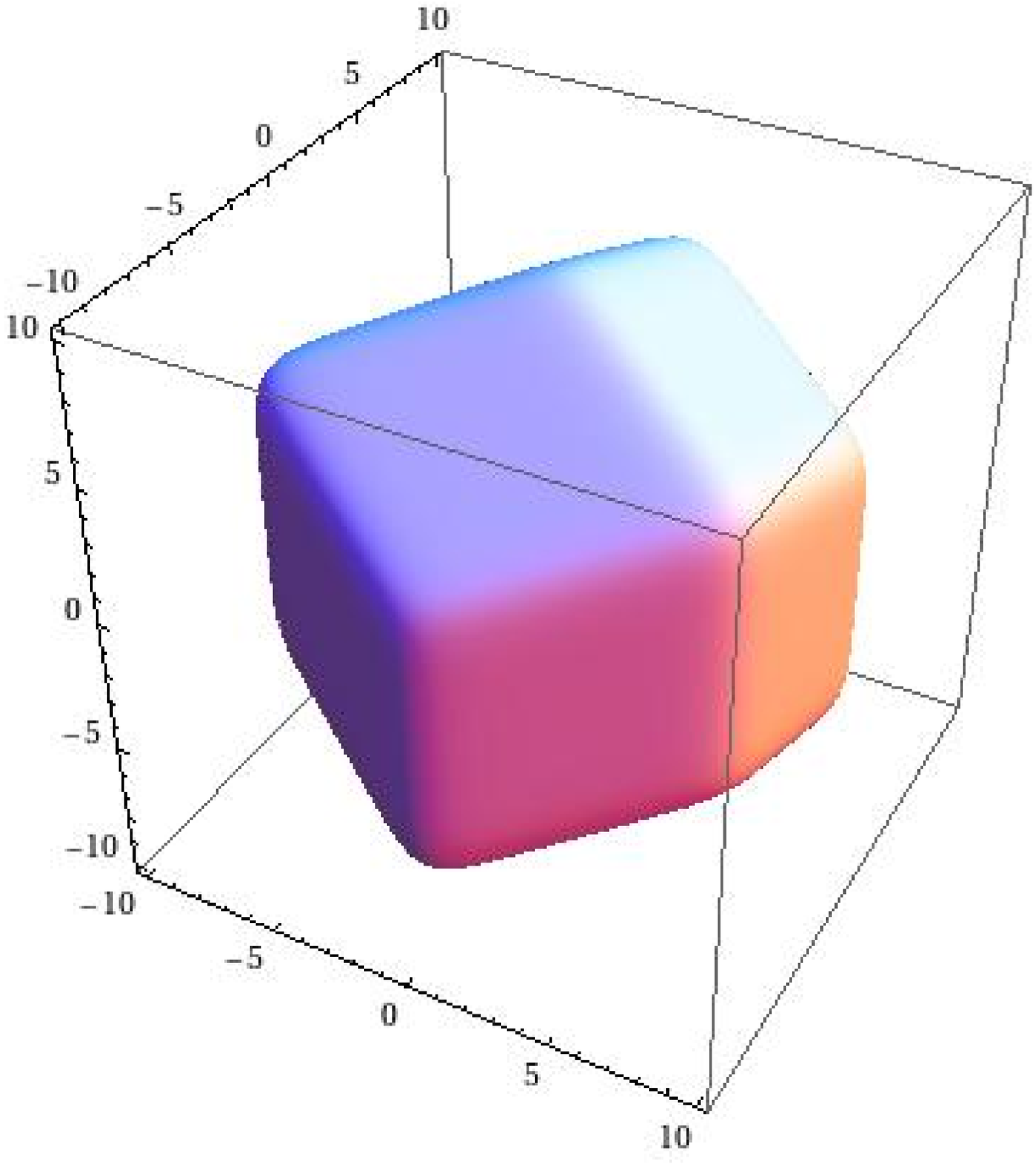,height = 3in}$\quad$\epsfig{file=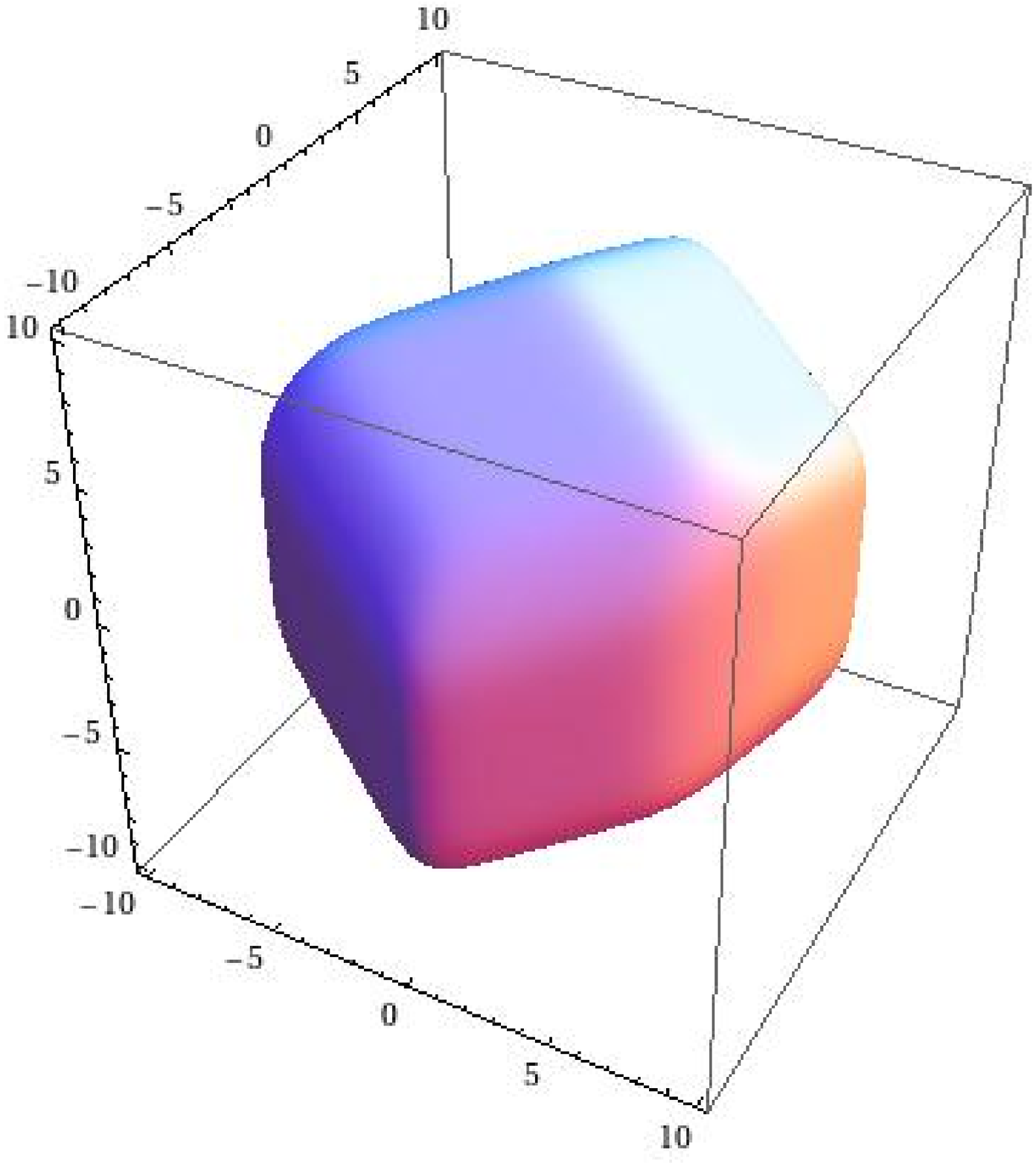,height=3in}

$ $

\epsfig{file=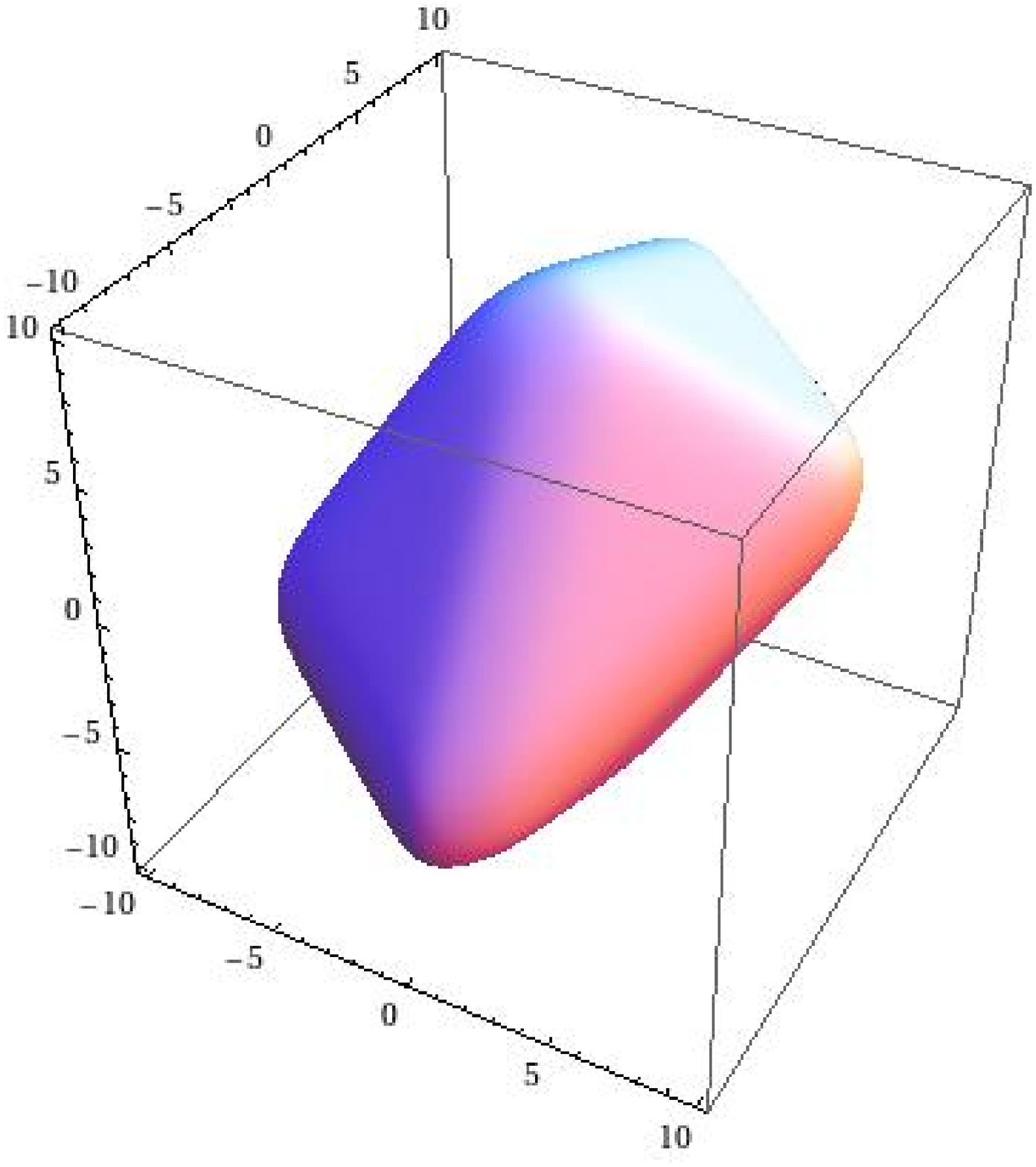,height=3in}$\quad$\epsfig{file=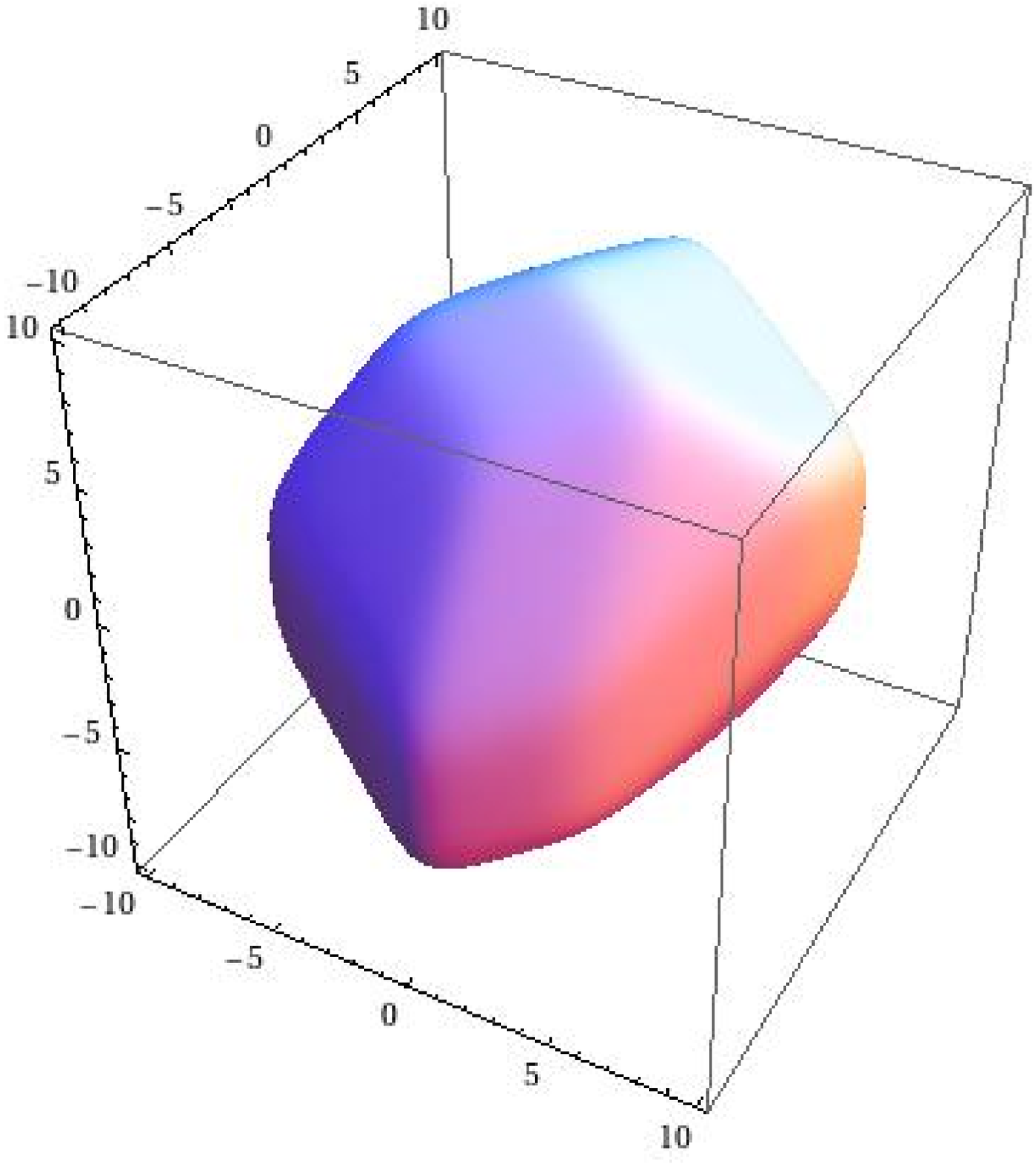,height = 3in}
\caption{The contour $\log\left(C_{1234}/m^2\right)=17$ with, clockwise from top left, $k=\8$, $1000i$, $100i$, $10i$.}\label{C4pic}
\end{figure}

We turn now to states of more than two particles. The two-particle case above was especially simple because there was only one exchange operation to be defined, but the broad approach will still apply. 
Consider first the case of three-particle states.
The goal is to define in a $\kappa$-covariant fashion two maps, 
\be \tau_{12},\,\,\tau_{23}\,\,: V_m\otimes V_m\otimes V_m \longrightarrow V_m\otimes V_m \otimes V_m\ee which generate a copy of the symmetric group $S_3$,
\be \tau_{12}^2 = \tau_{23}^2 = \id,\qquad \tau_{12} \,\tau_{23} \,\tau_{12} = \tau_{23}\, \tau_{12} \,\tau_{23},\label{S3def}\ee
and are such that in the limit $\kappa\rightarrow \8$ the map $\tau_{12}$ ($\tau_{23}$) just exchanges the factors $1$ and $2$ (respectively $2$ and $3$) in the tensor product.

With rapidity parameters as in section \ref{singlep}, let us write the basis states of
$V_m^{\otimes 3}$ as
\be \ket{r_0(z_1),r_1(z_1)} \otimes \ket{s_0(z_2),s_1(z_2)} \otimes \ket{t_0(z_3),t_1(z_3)}.\label{3P}\ee
The three-particle boost operator is then (cf \ref{2boost})
\bea N &=&  e^{-\frac{r_0}{2\kappa}-\frac{s_0}{2\kappa}}\pd{z_3} + e^{-\frac{r_0}{2\kappa} +\frac{t_0}{2\kappa}}\pd{z_2} + e^{\frac{s_0}{2\kappa} +\frac{t_0}{2\kappa}}\pd{z_1} \\
&=&{}  \left( \Cn z_1 + i\Sn z_1 \right)  \left( \Cn z_2 + i\Sn z_2 \right)\pd{z_3} \nn\\
 &&{}          + \left( \Cn z_1 + i\Sn z_1 \right)  \left( \Cn z_3 - i\Sn z_3 \right)\pd{z_2}\nn\\ 
 &&{}          + \left( \Cn z_2 - i\Sn z_2 \right)  \left( \Cn z_3 - i\Sn z_3 \right)\pd{z_1} \label{3boost}\eea
and all functions in the kernel of $N$ can be expressed in terms of 
\bea \mu_{12} &=& -\half \log \frac{1}{k}\left( \Dn z_1 - k\Cn z_1\right)\left( \Dn z_1 - i k \Sn z_1\right)\\
         && -\half \log \frac{1}{k}\left(  \Dn z_2 - k\Cn z_2\right)\left( \Dn z_2 + i k \Sn z_2\right)
          +\half \log \left( 1-\frac{1}{k^2} \right) \nn\\
 \mu_{23} &=& -\half \log \frac{1}{k}\left( \Dn z_2 - k\Cn z_2\right)\left( \Dn z_2 - i k \Sn z_2\right)\\
         && -\half \log \frac{1}{k}\left(  \Dn z_3 - k\Cn z_3\right)\left( \Dn z_3 + i k \Sn z_3\right)
          +\half \log \left( 1-\frac{1}{k^2} \right) \nn.\eea
In particular, it is a lengthy but straightforward exercise to verify that the value $C_{123}$ of the  Casimir $(\Delta \otimes \id)\Delta \cas$ on the state (\ref{3P}) is given by
\bea \frac{C_{123}}{m^2} &=& \left(1-\frac{1}{k^2}\right)^2 \left( e^{2\mu_{12}+2\mu_{23}} + e^{-2\mu_{12}-2\mu_{23}} \right) \nn\\
&& - \frac{1}{k^2} \left(1-\frac{1}{k^2}\right) \left( e^{2\mu_{12}-2\mu_{23}} +
e^{-2\mu_{12}+2\mu_{23}} \right) \nn\\
&& +\left(1-\frac{1}{k^2}\right) \left(1-\frac{2}{k^2}\right) 
 \left( e^{2\mu_{12}} + e^{-2\mu_{12}} + e^{2\mu_{23}} + e^{-2\mu_{23}}  \right)\nn \\
&& + 3\left(1-\frac{1}{k^2}\right)^2 + \frac{1}{k^4}. \label{Cas3}\eea

Now, as in the two-particle case, it is possible to pass from $(z_1,z_2,z_3)$ to new coordinates
\be  (\mu_{12}, \mu_{23}, \phi) \ee
on the space of basis states of $V_m^{\otimes 3}$ in such a way that the total momentum $P_\mu$ of a state depends on $\mu_{12}$, $\mu_{23}$ \emph{only} through the value of the Casimir. (We can, for example, pick $\phi= P_1$ or pick $\phi$ to be the coordinate such that $N=\pd \phi$.)
In such coordinates, let $I$ be any map of the form
\be  I: (\mu_{12}, \mu_{23}, \phi) \mapsto  (\tilde\mu_{12}(\mu_{12},\mu_{23}),\,\, \tilde\mu_{23}(\mu_{12},\mu_{23}),\,\, \phi) \ee
with the property that $I$ preserves $C_{123}$. Then $I$ defines an intertwiner of $\kappa$-Poincar\'e. 

This is true essentially by construction: $I$ commutes with boosts because the boost operator $N$ preserves $\mu_{12},\mu_{23}$, and the demand we made on the coordinate $\phi$, together with the condition that $I$ preserves $C_{123}$, ensures that $I$ preserves the momentum of states.

We must therefore find maps which are symmetries of the contours of the Casimir function. In contrast to the two-particle case, in which $\mu\mapsto -\mu$ was the only possibility, here there is considerable freedom. The contours of $C_{123}$ in the plane of $\{\mu_{12},\mu_{23}\}$ are illustrated in figure \ref{C3pic}, for various values of $k$. 
We have chosen to draw the $\mu_{12}$ and $\mu_{23}$ axes $120^\circ$ apart. In the limit of vanishing deformation the contours are then ``rounded'' hexagons, and are, as one would expect, preserved by the group $A_2 (\cong S_3)$ of rigid Euclidean reflections. Since \be\mu_{12}\rightarrow\half(z_1-z_2), \qquad \mu_{23} \rightarrow \half(z_2-z_3),\qquad\text{as}\quad k\rightarrow \8\ee 
these reflections are indeed nothing but the exchange of particle rapidities: for example, reflection in the simple root $\alpha_1$ is $\mu_{12}\mapsto -\mu_{12}; \mu_{23}\mapsto \mu_{12} + \mu_{23}$, which is just $z_1\leftrightarrow z_2$. 

As $\kappa$ decreases, this rigid $A_2$ reflection symmetry of the contours is lost -- which is no surprise: one would not expect the action of $S_3$ we seek to be linear in these variables. The important point, however, is that the contours are still topologically circles. It is in fact clear that this must be so from inspection of (\ref{Cas3}), in which, recall, $k^2<0$. Consequently, there certainly exist (suitably continuous) realizations of $S_3$.

One realization of $S_3$ is obtained by projecting any given contour onto the unit circle along rays through the origin (possible since the contours are star-shaped about the origin), letting an element of $S_3$ act on the circle by rigid reflections of the plane, and then projecting back.
This has the correct limiting behaviour by construction, and also respects the symmetries $(\mu_{12},\mu_{23}) \mapsto (\mu_{23}, \mu_{12})$ and $(\mu_{12},\mu_{23}) \mapsto (-\mu_{12},-\mu_{23})$ apparent in figure \ref{C3pic}. But it is not clear that it is the preferred way for $S_3$ to act. There are, in particular, many ways to identify a given contour with the unit circle, and we expect that a preferred identification is picked out by some additional criteria we have not introduced. 

Not all extra requirements on $\tau_{12}$, $\tau_{23}$ are consistent of course. It might, for example, appear natural to demand that $\tau_{12}$ ($\tau_{23}$) preserve the value of the corresponding two-particle Casimir $C_{12}$ ($C_{23}$). But the unique maps with \emph{these} properties which also preserve $C_{123}$ are, respectively,
\be \mu_{12} \mapsto -\mu_{12} ; \quad \mu_{23} \mapsto\mu_{23}+\log\left(\frac{(k^2 - 1)e^{\mu_{12}} -1}{k^2 - 1 - e^{\mu_{12}}}\right) \ee
and
\be \mu_{12} \mapsto \mu_{12} + \log\left(\frac{(k^2-1)e^{\mu_{23}} -1}{k^2 -1 -e^{\mu_{23}}}\right) ; \quad \mu_{23} \mapsto -\mu_{23} \ee
and for any finite $k\in i\mathbb R$ they fail to obey the braid relation $\tau_{12}\tau_{23}\tau_{12} = \tau_{23}\tau_{12}\tau_{23}$.

\subsubsection*{Four particles}
Finally, let us discuss the 4-particle case briefly. Apart from the extra dimension the situation is entirely similar. It may be verified by direct computation that the total Casimir can be expressed in terms of the $\mu_{i,i+1}$ as follows.
\bea \frac{C_{1234}}{m^2} &=&{}+\left(1-\frac{1}{k^2}\right)^3 
       \left(  e^{2\mu_{12} + 2\mu_{23} + 2\mu_{34}} + e^{-2\mu_{12} - 2\mu_{23} - 2\mu_{34}}\right)\label{Cas4}\\
 &&{} -\frac{1}{k^2} \left(1-\frac{1}{k^2}\right)^2
       \left( e^{-2\mu_{12} + 2\mu_{23} +2\mu_{34}} + e^{2\mu_{12} + 2\mu_{23}-2\mu_{34}}
             +e^{-2\mu_{12} -2\mu_{23} +2\mu_{34}} + e^{2\mu_{12} -2\mu_{23} - 2\mu_{34}} \right)\nn\\
 &&{} + \frac{1}{k^4} \left(1-\frac{1}{k^2}\right) 
       \left( e^{2\mu_{12} - 2\mu_{23} + 2\mu_{34}} + e^{-2\mu_{12} + 2\mu_{23}-2\mu_{34}}\right)\nn\\
 &&{} +\left(1-\frac{1}{k^2}\right)^2 \left(1-\frac{2}{k^2} \right) 
       \left( e^{2\mu_{12} +2\mu_{23}} +  e^{-2\mu_{12} -2\mu_{23}} 
             + e^{2\mu_{23} +2\mu_{34}} +  e^{-2\mu_{23} -2\mu_{34}}\right) \nn\\
  &&{} - \frac{1}{k^2}\left(1-\frac{1}{k^2}\right) \left(1-\frac{2}{k^2} \right)
       \left( e^{2\mu_{12} -2\mu_{23}} +  e^{-2\mu_{12} +2\mu_{23}} 
             + e^{2\mu_{23} -2\mu_{34}} +  e^{-2\mu_{23} +2\mu_{34}}\right) \nn\\
 &&{} - \frac{2}{k^2} \left(1-\frac{1}{k^2}\right)^2
       \left( e^{2\mu_{12} +2\mu_{34}} +  e^{-2\mu_{12} +2\mu_{34}} 
             + e^{2\mu_{12} -2\mu_{34}} +  e^{-2\mu_{12} -2\mu_{34}}\right) \nn\\
&&{} +\left(1-\frac{1}{k^2}\right) \left(1-\frac{2}{k^2}\right)^2
      \left( e^{2\mu_{12}} +  e^{-2\mu_{12}} + e^{2\mu_{23}} +  e^{-2\mu_{23}} + e^{2\mu_{34}} +  e^{-2\mu_{34}} \right)  \nn\\
&&{} +4 \left(1-\frac{1}{k^2}\right)^3 + \frac{4}{k^4} \left(1-\frac{1}{k^2}\right) \nn,\eea
Figure \ref{C4pic} illustrates the contours of this function. Once more, with the $A_3(\cong S_4)$ root system in mind, the plot is drawn with the $\mu_{23}$ axis at $120^\circ$ to both the $\mu_{12}$ and $\mu_{34}$ axes, and the latter pair at right angles. One sees that for large $k$ the contours are preserved by the rigid Euclidean reflections of $A_3$ and that for all $k\in i\mathbb R$, $k>1$, they are topologically 2-spheres. 

\section{Conclusions and open questions}\label{conc}
In this paper we examined tensor products of the spin-zero representation of $\kappa$-Poincar\'e in 1+1 dimensions. We showed that for two-particle states there exists a unique covariant definition of particle exchange -- that is, a unique non-trivial intertwiner. 
For states of three and four particles our result is that intertwining actions of the symmetric group exist, but the uniqueness of the two-particle case is lost. The key point is that, in the tensor product of $n\leq 4$ particles, the set of irreducible components with any given common value of the Casimir has the topology of the sphere $\mathbb S^{n-2}$. Whereas there is only one non-trivial way for $S_2$ to act on $\mathbb S^0=\{\pm1\}$, for $n\geq 3$ there are many actions of $S_{n}\cong A_{n-1}$ on $\mathbb S^{n-2}$ with the correct $\kappa\rightarrow \8$ limit.

It seems reasonable to expect that the pattern of spheres persists for $n>4$ particles, but further work is needed to verify this. We arrived at the expressions (\ref{Cas3}) and (\ref{Cas4}) for the three- and four-particle Casimirs by somewhat lengthy direct computation. There should be a more insightful approach which would yield the general expression for $C_{1\dots n}$. We anticipate that there is some preferred choice of variables in which these expressions simplify and it becomes clearer how the $\tau_{ii+1}$ should act. Finding these variables amounts to finding the natural labelling of many-particle states in $\kappa$-deformed theories, and is in a sense the central challenge for future work in this direction.

\vspace{1cm}
\emph{Acknowledgements} C.Y. is grateful to the Leverhulme trust for financial support. R.Z. is supported by an EPSRC postdoctoral fellowship.

\appendix
\section{Perturbative results for three particles in 3+1 dimensions}
We follow the notation and conventions (in particular, the use of the bicrossproduct basis) of \cite{IT1}. By methods entirely analogous to the two-particle case given in that paper, it may be verified that, to order $\frac{1}{\kappa^2}$, there is a one-parameter family of maps $(\tau_{12},\tau_{23})$ that obey the defining relations of $S_3$, as in (\ref{S3def}) above, commute with the action of $\kappa$-Poincar\'e, and preserve the masses of the individual tensor factors. These maps, parameterized by $a\in \mathbb R$, are
\bea
\tau_{12} &:& \ket{r_0, r_i} \otimes \ket{s_0, s_i}\otimes \ket{t_0 ,t_i} \longrightarrow\nn \\
&& \left| s_0 +\frac{1}{\kappa} \vec r \cdot \vec s + \frac{1}{\kappa^2} \left [ a \left (- r_0 \vec s \cdot \vec t + t_0 \vec r \cdot \vec s \right )
-\frac{1}{2} r_0 \vec s \cdot \vec s + \frac{1}{2} s_0 \vec r \cdot \vec r + \frac{1}{2} r_0 \vec r \cdot \vec s  + \frac{1}{2} s_0 \vec r \cdot \vec s \right ] \right .\, , \nn \\
&& 
\,\,s_i + \frac{1}{\kappa} s_0 r_i + \frac{1}{\kappa^2} \left [ \frac{1}{2}  r_i \vec r \cdot \vec s - \frac{1}{2} s_i \vec r \cdot \vec s -\frac{1}{2} s_i r_0 s_0  -\frac{1}{2} r_i s_0^2 + \frac{1}{2} r_0 r_i s_0 \right . \nonumber \\
&& \qquad \qquad\qquad \qquad\qquad \qquad \left .+ a
\left ( - r_i \vec s \cdot \vec t + t_i \vec r \cdot \vec s + r_i s_0 t_0- t_i r_0 s_0 \right )\right] \bigg>  \nonumber \\
&\otimes& \bigg| r_0 - \frac{1}{\kappa} \vec r \cdot \vec s + \frac{1}{\kappa^2} \left [a \left ( s_0 \vec r \cdot \vec t - t_0 \vec r \cdot \vec s \right ) + \frac{1}{2} r_0 \vec s \cdot \vec s - \frac{1}{2} s_0 \vec r \cdot \vec r - \frac{1}{2} r_0 \vec r \cdot \vec s - \frac{1}{2} s_0 \vec r \cdot \vec s \right ]\, , \nonumber \\
&& \,\,r_i - \frac{1}{\kappa} r_0 s_i + \frac{1}{\kappa^2} \left [ \frac{1}{2} r_i \vec r \cdot \vec s + \frac{1}{2} s_i \vec r \cdot \vec s - \frac{1}{2} r_0 r_i s_0
-\frac{1}{2}s_i r_0  s_0 + \frac{1}{2} s_i r_0^2 \right . \nonumber \\
&& \qquad \qquad\qquad \qquad\qquad \qquad \left .+ a \left ( s_i \vec r \cdot \vec t - t_i \vec r \cdot \vec s  -r_0 s_i t_0 + t_i  r_0 s_0 \right ) \right ]\bigg>  \nonumber \\
&\otimes& \bigg| t_0 + \frac{a}{\kappa^2} \left (r_0 \vec s \cdot \vec t - s_0 \vec r \cdot \vec t \right ) \, , \nonumber \\
&& \,\,\left . t_i + \frac{a}{\kappa^2} \left (r_i \vec s \cdot \vec t - s_i \vec r \cdot \vec t -r_i s_0 t_0 + r_0 s_i t_0 \right ) \right )\bigg>
\eea
\bea
\tau_{23}&:&  \ket{r_0, r_i} \otimes \ket{s_0, s_i}\otimes \ket{t_0 ,t_i} \longrightarrow\nn \\
&& \left| r_0 + \frac{a-1/3}{\kappa^2} \left ( s_0 \vec r \cdot \vec t - t_0 \vec r \cdot \vec s  \right ) \, ,  \right .\nonumber \\
&& \,\,r_i + \frac{a-1/3}{\kappa^2} \left ( s_i \vec r \cdot \vec t - t_i \vec r \cdot \vec s +t_i r_0 s_0 - s_i r_0 t_0 \right )\bigg> \nonumber \\ 
&\otimes& \bigg|t_0 + \frac{1}{\kappa} \vec s \cdot \vec t + \frac{1}{\kappa^2} \left [ \left (a -\frac{1}{3} \right ) \left ( r_0 \vec s \cdot \vec t - s_0 \vec r \cdot \vec t \right )
+\frac{1}{2} t_0 \vec s \cdot \vec t + \frac{1}{2} s_0 \vec s \cdot \vec t - \frac{1}{2} s_0 \vec t \cdot \vec t + \frac{1}{2} t_0 \vec s \cdot \vec s \right ] \, , \nonumber \\
&& \,\,t_i + \frac{1}{\kappa} t_0 s_i + \frac{1}{\kappa^2} \left [ \left (a -\frac{1}{3} \right ) \left ( r_i \vec s \cdot \vec t  - s_i \vec r \cdot \vec t +s_i r_0 t_0 -r_i s_0 t_0   \right ) \right . \nonumber \\
&& \qquad \qquad\qquad \qquad\qquad \qquad \left . + \frac{1}{2} s_i \vec s \cdot \vec t - \frac{1}{2} t_i \vec s \cdot \vec t +\frac{1}{2} s_0 s_i t_0-\frac{1}{2} s_i t_0^2 - \frac{1}{2} s_0 t_i t_0 \right ] \bigg> \nonumber \\
&\otimes&\bigg| s_0 - \frac{1}{\kappa} \vec s \cdot \vec t + \frac {1}{\kappa^2} \left [\left (a -\frac{1}{3} \right ) \left ( t_0 \vec r \cdot \vec s - r_0 \vec s \cdot \vec t   \right ) + \frac{1}{2} s_0 \vec t \cdot \vec t - \frac{1}{2} s_0 \vec s \cdot \vec t - \frac{1}{2} t_0 \vec s \cdot \vec t - \frac{1}{2} t_0 \vec s \cdot \vec s \right ]\, , \nonumber \\
&& \,\,s_i - \frac{1}{\kappa} s_0 t_i + \frac{1}{\kappa^2} \left [\left (a -\frac{1}{3} \right ) \left (t_i \vec r \cdot \vec s -r_i \vec s \cdot \vec t  + r_i s_0 t_0 - t_i r_0 s_0    \right ) \right . \nonumber \\
&& \qquad \qquad\qquad \qquad\qquad \qquad \left . \left .+ \frac{1}{2} s_i \vec s \cdot \vec t + \frac{1}{2} t_i \vec s \cdot \vec t
-\frac{1}{2} s_0 s_i t_0+\frac{1}{2} t_i s_0^2 -\frac{1}{2} s_0 t_i t_0 \right ]  \right ) \bigg>.
\eea
Observe that, writing $\tau$ for the two-particle intertwiner of \cite{IT1}, one has $\tau_{12}=\tau\otimes 1$ when $a=0$, while if $a=\frac{1}{3}$ then $\tau_{23}=1\otimes \tau$. But for no value of $a$ is it true that \emph{every} $\tau_{ii+1}$ acts non-trivially only on the i,i+1st tensor factors -- which accords with the results of section \ref{manyparticles} above.

\end{document}